\documentclass[10pt,aps,prb,twocolumn,preprintnumbers,amsmath,amssymb,floatfix,showpacs,citeautoscript,superscriptaddress]{revtex4-2} %
\usepackage[latin1]{inputenc}
\usepackage[T1]{fontenc}
\usepackage[english]{babel}
\usepackage[pdftex]{graphicx}
\usepackage{ulem}
\usepackage{amsmath}
\usepackage{times}
\usepackage{subfigure}
\usepackage[usenames,dvipsnames,svgnames,table]{xcolor}
\usepackage[colorlinks,plainpages=false,linkcolor=blue,urlcolor=blue,citecolor=blue,pdfpagemode=UseNone]{hyperref}
\usepackage{svg}
\newcommand{\mub}{$\mu_{\rm B}$}
\newcommand{\ef}{$E_{\rm F}$}

\renewcommand{\AA}{\text{\r{A}}}

\newcommand\Vek[1]{\vec{#1}}

\newcommand\muB{\mu_{\text{B}}}

\begin{document}

\title
{
\boldmath
Engineering the band structure of few-layer graphene by S-doping: from  linear dispersion to impurity flat bands%
}

\author{Armin Sahinovic}
\affiliation{Department of Physics and Center for Nanointegration (CENIDE), Universit\"at Duisburg-Essen, Lotharstr.~1, 47057 Duisburg, Germany}
\author{Rossitza Pentcheva}
\email{rossitza.pentcheva@uni-due.de}
\affiliation{Department of Physics and Center for Nanointegration (CENIDE), Universit\"at Duisburg-Essen, Lotharstr.~1, 47057 Duisburg, Germany}

\date{\today}

\begin{abstract}
    Motivated by the technological relevance of S-doped few-layer graphene (FLG) in battery applications and in the oxygen reduction reaction, 
    we systematically explore the effect of basal plane S-doping on the electronic properties of mono-, bi-, and four-layer graphene, using first-principles calculations with van der Waals corrections. 
    In the monolayer we find a variety of effects ranging from a sustained Dirac cone with localized impurity bands away from the Fermi level for thiophenic doping (2V1S) to a band gap opening of 0.4~eV and impurity flat bands close to the Fermi-level for graphitic doping (1V1S) and an additional $n$-type doping together with spin-polarization for thiopyranic doping (4V3S). Incorporation in FLG leads to modification of the Dirac cone into a set of hyperbolic touching bands in 2V1S; reduction (bilayer) and closing of the band gap  with additional hyperbolic touching bands in conjunction with an impurity flat band at the Fermi level in 1V1S and 4V3S and a reduction of spin polarization in the latter. Overall, S-doping enables design of the band structure and tuning the electronic behavior of FLG from metallic to insulating and from linear dispersion to impurity flat bands that makes S-doped FLG a promising material for versatile technological applications. 
\end{abstract}

\maketitle

\section{Introduction}

Since the initial discovery of graphene~\cite{novoselov_electric_2004} a substantial effort has focused on tailoring the exceptional properties of 2D materials for device applications~\cite{ferrari_science_2015} by defects, doping, adsorbates, interaction with the substrate or application of a gate voltage~\cite{liu_chemical_2011, ferrighi_boron-doped_2015}. 
Vacancies may occur naturally, or be artificially introduced using chemical processing or electron, ion and laser irradiation~\cite{Schleberger_2018, Atomistic_Scale_Simulations_2016, COMPAGNINI_ion_irradtion_2009, Russo2013_porous_graphene, Qin2024_nanopores, Sun_nanopores_2025, laser_nanoprocessing_2023, Banhart:2011}.
Unsaturated bonds either at edge sites or at vacancies promote different incorporation mechanisms into the graphene structure.
For example, N-dopants are preferentially found in a pyridinic and pyrrolic configuration~\cite{Kattel2012}. Intensive research has been focused on dopants like nitrogen or boron that modify the number of charge carriers leading to $n$- and $p$-doping, respectively~\cite{ferrighi_boron-doped_2015, Casolo_band_engineering_BN_2011, Li_2015_n_doping}. 

In contrast to N or B doping, the electronic structure of S-doped graphene is much less explored despite its technological relevance for applications in energy storage such as supercapacitors and batteries~\cite{Yang2021_graphene_battery_sulfur, wu_bottom-up_2017,parveen_simultaneous_2015, Garcia:2008:1546-1955:2221, ADJIZIAN2013256, Qiu_Thiophene:23} or as a metal-free-cathode catalyst in the oxygen reduction reaction~\cite{S-doped_graphene_basal_plane:2026, Yin_2023_Thiophene_ORR, higgins_development_2014, priyadarsini_effects_2021, yang_sulfur-doped_2012, Jeon_2013_sulfurized_edge}. Moreover, it was shown that S-doping promotes the stronger adsorption of Pt in graphene~\cite{higgins_development_2014}, improving the ORR activity 
compared to commercially available Pt/C cathodes~\cite{park_sulfur-doped_2013, higgins_development_2014}.
S-doping introduces two additional electrons and has a similar electronegativity to C~\cite{higgins_development_2014, Jeon_2013_sulfurized_edge, tucek_sulfur_2016}, which is expected to lead to less localized effects compared to N-doping, which introduces one additional electron~\cite{Chaban_N_doped_graphene_stabililty_2015}.
Experimental results indicate incorporation of S-dopants in the basal plane~\cite{Bianco_Defect_healing_sulfurization_2024,S-doped_graphene_basal_plane:2026} and a homogeneous distribution in the graphene sheet~\cite{Fortugno_Sahinovic_S-graphene_2025}. 
Different basal-plane incorporation geometries were proposed, e.g., graphitic, thiopyranic or thiophenic~\cite{S-doped_graphene_basal_plane:2026, Wang-S-doped_graphene_Li:2019, Fortugno_Sahinovic_S-graphene_2025, Qiu_Thiophene:23, higgins_development_2014, Bianco_Defect_healing_sulfurization_2024, Lopez-Urias_edge_chemistry_2020}, the latter is promising to increase catalytic performance in the oxygen reduction reaction~\cite{Yin_2023_Thiophene_ORR}. The different dopant configurations can have distinct effect on the electronic and transport properties requiring a detailed understanding of their influence. 
A further unsettled question refers to the role of S in the emergence or quenching of spin-polarization in graphene: while some studies report spin-polarization in S-doped graphene~\cite{tucek_sulfur_2016}, others suggest that S quenches the spin polarization of defective graphene~\cite{zhu_magnetic_2016}. 

Scalable plasma-based growth~\cite{Levchenko_plasma_gr_growth_review_2016, Fortugno_Sahinovic_S-graphene_2025, Umut_Armin_PECVD_2025} or liquid exfoliation~\cite{Pope_2013} often leads to few-layer graphene for applications in opto-electronics and energy storage devices~\cite{Ngo_CuO_rGO_photodetector_2024, Bacher_graphene_leds_2024, han_ultrasoft_2020}.  
Multilayer graphene, in particular twisted \cite{cao_unconventional_2018, tian_evidence_2023, Zhang2025_twist} and rhombohedrally stacked~\cite{Munoz_rhombohedral_2013, Zhou2021_rhomb_graphene_supercond}, has attracted a lot of interest due to the correlated physics and emergence of flat bands and van Hove singularities. 
Bernal-stacked multilayer graphene exhibits touching hyperbolic bands at the Fermi level~\cite{mak_evolution_2010, ohta_controlling_2006, partoens_graphene_2006, bostwick_symmetry_2007, han_ultrasoft_2020, crowther_strong_2012, rokni_layer-by-layer_2017, Ziegler_screening_FLG_11} and can also host a broad set of intriguing phases, e.g., as a bilayer with applied displacement electric field~\cite{Seiler2024, PRLSeiler_2024, Seiler2022}. 
Flat bands can also be generated by impurities in graphene and have been proposed to host strong correlations and superconductivity~\cite{Marsal_enhanced_quantum_metric_2024, vanpoppelen2026disorderinducedsuperconductivitygraphene}.

While the effect of doping and defects has been studied thoroughly in single layer graphene for a series of dopants~\cite{ferrighi_boron-doped_2015, Friedrich_magn_b_doped_graphene_20, lee_tunable_2019, Alsaati2024, Wei_N-doped_Gr_09, Zhou_N_Al_doped, Bie_N_doped_Gr_Photocatal_21, cuxart_spatial_2023, Denis_Theo_s_n_doped_Gr_14}, 
it is unclear how the interlayer interactions in few-layer graphene used in scalable device fabrication modifies the electronic properties of the doped system.
To gain insight into the mechanism of S-incorporation and its influence on the electronic properties of monolayer, bilayer and 4L-FLG, we carried out density functional theory (DFT) calculations including van der Waals corrections. As a starting point we address the energetic stability and electronic properties of vacancies in graphene and FLG (Section~\ref{Sec:Vac}). Subsequently, the stability and structure of a wide range of S-doping configurations and an S-concentration varying between 2.00-6.25~\% in the monolayer, bilayer and 4L-FLG is explored in Section~\ref{Sec:en_S_doping}. Lastly, we analyze the electronic structure of the S-doped monolayer and multilayer graphene in Section~\ref{Sec:Bands}. 
We demonstrate a wide set of intriguing effects,
ranging from preserved Dirac cones, e.g., for thiophenic incorporation to Dirac cone splitting and band gap opening in conjunction with the emergence of impurity flat bands for graphitic incorporation. Moreover, spin polarization is predicted for thiopyranic S-doping 
with important distinctions concerning spatial localization to the one emerging due to vacancies. Remarkably, multilayer configurations allow one to superimpose the graphene/graphite-like band structure of the pristine layers on the band structure of the defect layer.
The results demonstrate that targeted S-doping can be used to tune graphene for various purposes, such as catalysis, energy storage and spintronics.

\section{Methods and Computational Details}\label{Sec:methods}
We performed first-principles simulations on pristine and sulfur-doped graphene and 
FLG with the projected augmented wave (PAW) method in the framework of the Vienna Ab initio Simulation Package (VASP)~\cite{USPP-PAW:99, PAW:94}, 
using the generalized gradient approximation as parameterized by Perdew, Burke, and Ernzerhof~\cite{PeBu96}. 
For accurate description of the long-range van der Waals 
interactions we use the Grimme DFT-D3 correction with Becke-Johnson damping ~\cite{Grimme-D3:10, Grimme-Damping:11}. The wave functions were expanded into plane waves up to a cutoff energy of 400 eV.
A Gaussian smearing of 0.075 eV was applied to the Kohn-Sham states. 
The density of states was calculated using the tetrahedron method with Bl\"ochl corrections~\cite{tetrahedron_bloechl_94}.
To minimize the interaction of dopants in neighboring unit cells and avoid the downfolding of K and K' to $\Gamma$~\cite{Casolo_band_engineering_BN_2011} we considered a 
$5\times5$ supercell for the monolayer and multilayer graphene systems. 
The integration over the Brillouin zone was performed using
a $5 \times 5 \times 1$  and a $24 \times 24 \times 1$ $\Gamma$-centered $\Vek{k}$-point grid for relaxation and density of states, respectively. 
The systems were separated by 20~\AA\ of vacuum to avoid spurious interaction with their periodic images and the positions of all atoms were fully relaxed until the forces were below $0.01$~eV\AA$^{-1}$.  
Band unfolding into the primitive cell of graphene was carried out using the method introduced by Popescu and Zunger as implemented in VASPKIT~\cite{VASPKIT, Popescu_Zunger_2010, Popescu_Zunger_2012}. 

Besides graphitic incorporation, we have explored additional binding configurations in defective graphene which we denote by $N_\text{V}$V$N_\text{S}$S, where $N_\text{V}$ and $N_\text{S}$ represent the number of C-vacancies and S-dopants incorporated in the vacancy, respectively.
The vacancy formation energy is defined as:
\begin{equation}
\label{eq:vacancy}
    E_\text{f,vacancy} =  (E_{\text{$N_\text{V}$V$0$S:(FL)G}}-E_{\text{(FL)G}}+N_\text{V}E_\text{C})
\end{equation}
$E_{\text{$N_\text{V}$V$0$S:(FL)G}}$ denotes the total energy of the system with $N_\text{V}$ vacancies. The formation energy per vacancy is obtained by dividing  $E_\text{f,vacancy}$ by $N_\text{V}$.
The dopant binding energy to a vacancy is calculated as:
\begin{equation}
\label{eq:binding}
    E_\text{S-binding} = \frac{1}{N_\text{S}} (E_{\text{$N_\text{V}$V$N_\text{S}$S:(FL)G}}-E_{\text{$N_\text{V}$V$0$S:(FL)G}}-N_\text{S}E_\text{S})
\end{equation}
$E_{\text{$N_\text{V}$V$N_\text{S}$S:(FL)G}}$ and $E_{\text{$N_\text{V}$V$0$S:(FL)G}}$ denote the total energy of the doped $N_\text{V}$V$N_\text{S}$S system and the reference $N_\text{V}$V$0$S system, respectively, while $E_\text{S}$ is the reference energy per S atom of the S$_8$-ring structure as the most common phase of sulfur~\cite{Meyer1976-lw}. 
To determine the thermodynamic stability of different doping configurations we use  the the S-incorporation energy with respect to pristine (few-layer) graphene as: 
\begin{equation}
\label{eq:incorporation}
    E_\text{S-incorp.} = (E_{\text{$N_\text{V}$V$N_\text{S}$S:(FL)G}}-E_{\text{(FL)G}}+N_\text{V}E_\text{C}-N_\text{S}E_\text{S})\\
\end{equation}
where $E_{\text{(FL)G}}$ represents the total energy of the pristine system without defects.

\section{Vacancies in monolayer and few-layer graphene}
\label{Sec:Vac}

\begin{figure}[tb]
    \includegraphics[width=1\linewidth]{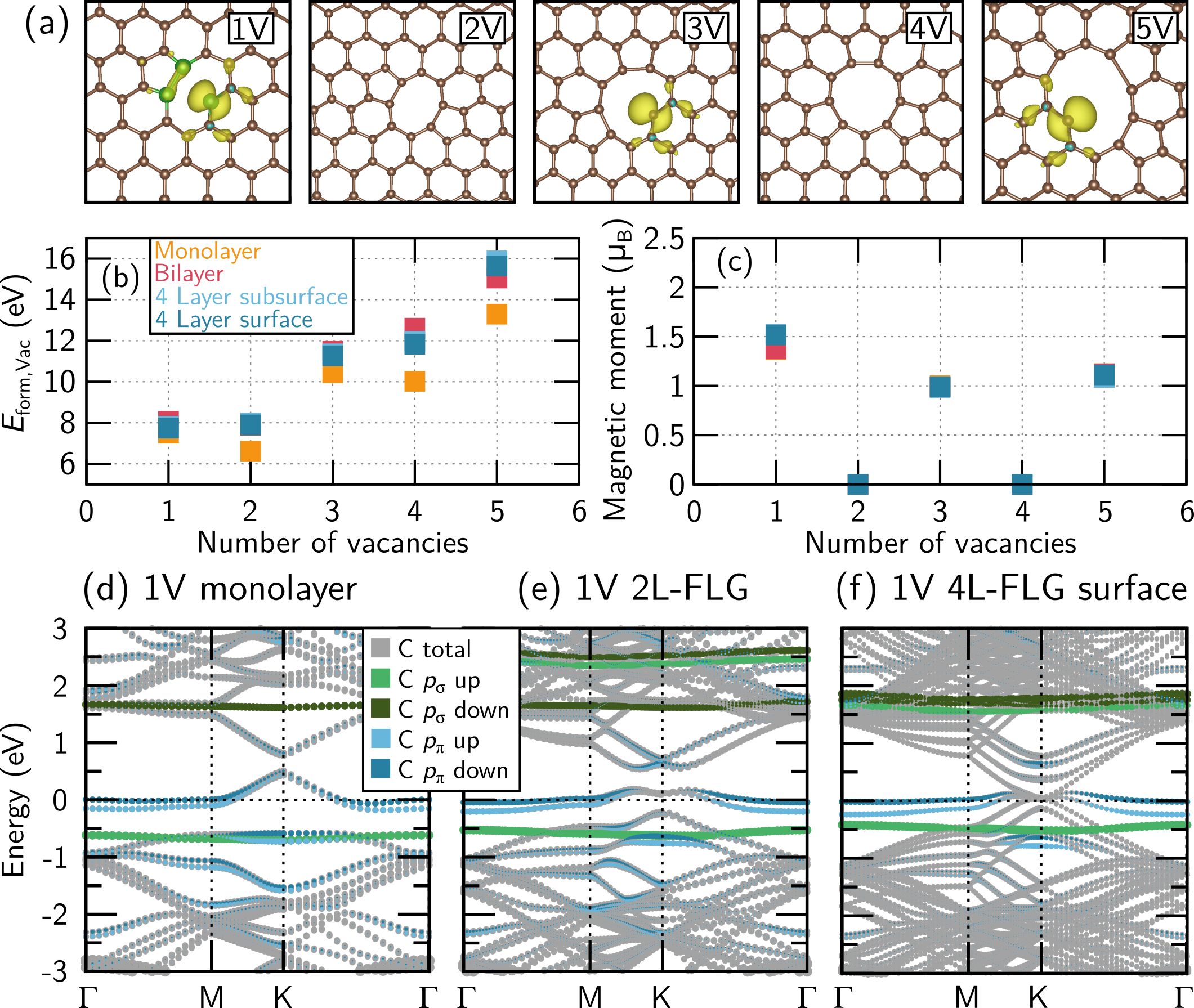}
    \caption{
    (a) Top view of the relaxed vacancy configurations $N$V, $N$ indicating the number of removed C-atoms. 
    The C-atoms are shown in a brown color. The nearest C- neighbors to the vacancy are marked in green.
    The yellow and cyan color depict the majority and minority spin-density, respectively. 
    (b) Vacancy formation energy and (c) total magnetic moment of the system with vacancies as a function of the number of removed C-atoms. 
    Orange, red, light blue and dark blue squares denote the vacancy in a monolayer, bilayer, subsurface and surface 4L-FLG, respectively.
    Element-, orbital- and spin-resolved band structure of the 1V single vacancy in (d) monolayer, (e) bilayer and (f) in a surface layer of 4L-FLG. 
    Total C contributions are shown in gray. 
    The majority/minority C $p_{\sigma}$ and C $p_{\pi}$ associated with the C-atoms marked in green are shown in green/dark green and blue/dark blue, respectively. 
    }
    \label{fig:structure_vacancy}
\end{figure}

Prior to exploring S-doping, we address the properties of different vacancies in graphene.
The formation of a mono-vacancy (1V) is associated with a high energy cost of $E_{\rm f} = 7.49$ eV (cf. Fig.~\ref{fig:structure_vacancy}), close to the experimental value of  $7.3 \pm 1$ eV in graphite~\cite{HENSON_graphite_vacancy}. 
Previous theoretical results range between $7$ and $8$~eV, depending on the method used~\cite{Ali_Muhammad_vacancy_concentration_graphene, Haldar_formation_energy_sv_graphene, Wang2013_formation_vac_graphene}. 
Interestingly, we find a minor enhancement of the vacancy formation energy in few-layer graphene: $7.73$ eV/ $7.87$ eV in the surface/subsurface layer of 4L-FLG, respectively.

Top views of the relaxed structures are shown in Fig.~\ref{fig:structure_vacancy}a. 
A significant distortion in 1V lowers the symmetry from $\text{D}_{3h}$ to $\text{C}_{2v}$ forming a 5-9 ring defect~\cite{Valencia_single_vac_DFT_mag_17}. The corresponding band structure of 1V in the monolayer is displayed in Fig.~\ref{fig:structure_vacancy}d. 
Removing a single C-atom breaks the sublattice symmetry which results in splitting of the Dirac cone and massive Dirac-fermions~\cite{RevModPhys_graphene_2009}, opening a band gap of 0.34~eV in 1V.
The valence band (VB) of predominantly C $p_{\pi}$ origin crosses the Fermi level around K. 
Its maximum (VBM) lies 0.52~eV above the Fermi level, indicating pronounced $p$-type doping related to the two-electron deficiency due to the missing C.  
The valence and conduction bands show nearly linear dispersion close to K and flatten towards $\Gamma$ and M. Away from K the VB exhibits a stronger spin splitting with the minority/majority spin band at/0.1~eV below the Fermi level. 
Additionally, two pairs of impurity flat bands of C $p_{\sigma}$-orbital character emerge, an occupied one at $-0.62$~eV and an empty one at 3.4 eV in the majority spin channel and two empty ones at 1.6 eV and 3.5 eV (not shown here) in the minority spin channel.   
The splitting of the $p_{\sigma}$-bands has been attributed to a Jahn-Teller effect~\cite{Poppvic_Jahn_teller_vacancy_graphene, Yazyev_2007, Nanda_2012}. 

As a result of the exchange splitting of bands, the single vacancy exhibits a substantial spin-polarization with a total magnetic moment of $1.5$~$\mu_\text{B}$, in agreement with  previous results~\cite{Poppvic_Jahn_teller_vacancy_graphene, ZHANG_tunable_magn_graphene_vac_20, Valencia_single_vac_DFT_mag_17, Septya_2021}. 
The largest contribution to the spin density of 1V, shown in Fig. \ref{fig:structure_vacancy}a, is localized at the exposed carbon atom in the 9-ring and has an in-plane $p_{\pi}$-orbital character, indicative of a dangling bond, with smaller contribution of the remaining C-sites surrounding the vacancy. 
The spin polarization of the single vacancy is interesting in view of spintronics applications, e.g., the $p_{\pi}$-band occupation and thus the magnetism can be tuned using a gate voltage~\cite{Han2014} or a scanning tunneling microscope tip~\cite{Zhang_STM_graphene_magnetism_16,ZHANG_tunable_magn_graphene_vac_20}. 

Before discussing the 1V incorporated in the multilayer, we elaborate briefly on the main features of the undoped graphene multilayer band structure. 
The ones of the undoped bilayer and 4L-FLG are displayed in Fig.~\ref{fig:bands_pristine_FLG} in Appendix~\ref{app:pristine-FLG}.  
Bernal stacking breaks the symmetry of the A and B sublattices in the layer~\footnote{In Bernal stacked graphene the atoms of the A sublattice of one layer lie above  the hexagon centers of the lower layer, while the atoms of the B sublattice are positioned directly above  the A sublattice of the layer underneath.} leading to 
 massive Dirac fermions~\cite{bostwick_symmetry_2007, mak_evolution_2010,electronic_structure_multilayer_2020}.
The linear dispersion of the Dirac cone is modified into two (four) pairs of valence and conduction hyperbolic bands, one pair (two pairs) touching at the Fermi-level and the remaining split apart in the bilayer (4L-FLG).
The splitting and dispersion varies with the layer number $N$ and position in the layer stacking sequence, with the surface layers exhibiting a smaller energy splitting. For further discussion of the band structure of multilayer Bernal stacked graphene, see Ref.~\cite{mak_evolution_2010}.  

We find a strong modification of the 1V bilayer band structure with two pairs of hyperbolic bands separated by a band gap of 0.42~eV (cf. Fig.~\ref{fig:structure_vacancy}e).
The minima of the two hyperbolic conduction bands are shifted away from K. In addition, the valence bands  of C $p_\pi$ character display a similar shift of the band maximum away from K with reduced $p$-type doping compared to the monolayer. Moreover, they are spin-split, in particular towards $\Gamma$ the majority band lies 0.17~eV below the Fermi level, whereas the minority band is at the Fermi level. 
On the other hand, the band structure of 1V  embedded in the surface layer of 4L-FLG, shown in Fig.~\ref{fig:structure_vacancy}f, exhibits four pairs of hyperbolic bands that are weakly spin split, with one pair touching just above the Fermi level. Similar to the bilayer case, the two valence bands exhibit reduced $p$-type doping with the maximum shifted from K and an enhanced spin splitting towards $\Gamma$ with the minority band at the Fermi level and the majority band 0.24~eV below the Fermi level. Furthermore, the occupied majority impurity flat band of C $p_\sigma$ character moves up in energy from -0.62~eV in the monolayer to -0.52~eV in the bilayer and  -0.42~eV (4L-FLG) and the empty flat bands are lowered in energy from $3.5$~eV in the monolayer to $2.4$~eV in the bilayer and $1.7$~eV in 4L-FLG. The reduction of energy between the occupied and empty flat bands reflects the increase of electronic screening in the multilayer graphene structure.

We now turn to the 2V vacancy which is known to exhibit good binding capabilities of molecules and transition metals~\cite{Sanyal2009, Krasheninnikov2009, defect_reactivity_graphene_2013}.
Its symmetric structure comprises 5-8-5 rings (cf. Fig.~\ref{fig:structure_vacancy}a). The formation energy of the 2V vacancy ($6.65$~eV) is lower compared to the mono-vacancy (cf. Fig.~\ref{fig:structure_vacancy}b) due to the saturation of dangling bonds. As a consequence, its spin-polarization is quenched. 
A further configuration with three 7-rings and 5-rings  commonly referred to in the literature as 555-777~\cite{Banhart:2011, Kim_2011_divacancy_kinetic_barrier, Kotakoski_point_defects_2011} is lower in energy but requires a high barrier to be formed  (cf. in Appendix~\ref{app:metastable}, Fig.~\ref{fig:meta-stable}c). Since it is also chemically less active, we focus here on the more reactive and experimentally commonly observed 2V, in line with previous studies~\cite{Banhart:2011, Kotakoski_point_defects_2011, Haldar2014, Sanyal2009, MATTUR_divacancy_adsorption_2025, Krasheninnikov2009}. 

We have also considered larger vacancy clusters which can be created using, e.g., electron, ion or laser irradiation or chemical synthesis~\cite{Schleberger_2018, Atomistic_Scale_Simulations_2016, COMPAGNINI_ion_irradtion_2009, Russo2013_porous_graphene, Qin2024_nanopores, Sun_nanopores_2025, laser_nanoprocessing_2023, Banhart:2011}.
The 3V configuration is comprised of two 5-rings and a 10-ring and has a formation energy of 10.44~eV (3.48~eV per site). 3V carries a magnetic moment of 1~$\mu_\text{B}$ with a spin density of in-plane $p_\pi$ orbital character localized at a C-site within the 10-ring connected to a distorted hexagon and smaller contributions at neighboring C-atoms (cf. Fig.~\ref{fig:structure_vacancy}a). 
4V features a central 9-ring surrounded by three 5-rings and six hexagons. Since the dangling bonds are saturated this configuration exhibits no spin-polarization. 
The formation energy of 4V in the monolayer is $10.02$~eV (2.51~eV per site), slightly lower than the one of 3V (10.44~eV).
Lastly, the 5V case consists of three 5-rings enclosing a large 11-ring. A single unsaturated bond in the large 11-ring results in a magnetic moment of 1.1~$\mu_\text{B}$, again with small contribution of neighboring sites. 
The 5V formation energy is $13.29$~eV (2.65~eV per site).
Notably, the spin-densities of the vacancy configurations considered here are strongly localized around the site with unsaturated bonds.
The emergence of spin-polarization due to vacancies can be partially explained by Lieb's second theorem on bipartite lattices~\cite{Liebs_theorem_1989}, which states that the ground state magnetization is proportional to the sublattice imbalance, $S=\frac{1}{2}(N_\text{B}-N_\text{A})$. However, the reconstructions taking place at vacancies result in a non-bipartite lattice which limits the application of Lieb's theorem. For example, the most stable configuration of 4V shown in Fig.~\ref{fig:structure_vacancy} exhibits a sublattice imbalance (one atom/three atoms removed from the A/B sublattice, respectively) but no spin-polarization emerges.

A monotonic increase of the total vacancy formation energy is observed with $N$, except for the 2V and 4V cases where the saturation of dangling bonds results in higher stability. 
On the other hand, the energy cost per removed C decreases with increasing vacancy size. This trend is consistent with previous molecular dynamics~\cite{Kotakoski_large_vacancies_2014} and DFT studies of larger vacancies~\cite{Haldar2014}.
Our investigation indicates that a single extended vacancy is energetically preferred compared to two smaller vacancies in proximity to each other, as shown in Appendix.~\ref{app:metastable}.

\begin{table}                                                                   
\centering                                                                       
    \begin{ruledtabular}                                                         
    \caption{Vacancy formation energy (cf. Eq.~\ref{eq:vacancy}) of 1V, 2V, 3V, 4V and 5V cases in eV.  }
        \begin{tabular}{lcccc}                                               
        Case & \multicolumn{4}{c}{  Vacancy formation energy [eV] } \\
        & 1L & 2L & 4L-subsurface & 4L-surface   \\
        \hline
        1V  &$7.49$ &$8.07$ &$7.83$ &$7.74$ \\
        2V  &$6.65$  &$7.88$    &$7.98$             &$7.90$          \\
        3V  &$10.44$  &$11.50$  &$11.36$           &$11.26$          \\
        4V  &$10.02$  &$12.61$  &$11.97$           &$11.83$          \\
        5V  &$13.29$  &$15.05$  &$15.89$           &$15.65$          \\
        \end{tabular}
        \label{tab:vac_form}
    \end{ruledtabular}
\end{table}

An overall increase in vacancy formation energy is found in the multilayer structures compared to the monolayer (cf. Fig.~\ref{fig:structure_vacancy}b and Table~\ref{tab:vac_form}).
In the bilayer we find a vacancy formation energy of 8.07 eV (1V), 7.87~eV (2V), 11.49~e (3V)V, 12.61~eV (4V), 15.05~eV (5V). 
Similar values are obtained for the vacancy formation energy in the surface (subsurface) layer of 4-FLG: 7.73~eV (7.82~eV) (1V), 7.89~eV (7.98~eV) (2V), 11.26~eV (11.36~eV) (3V), 11.83~eV (11.97~eV) (4V), 15.64~eV (15.89~eV) (5V), indicating that  the cost for vacancy formation in the subsurface layer is slightly higher.
The general increase in vacancy formation energy in the multilayer systems is attributed to interlayer interactions. 
On the other hand, the magnetic moments of vacancies in the multilayer are similar to the ones in the monolayer (cf. Fig~\ref{fig:structure_vacancy}c).

\section{Energetic Stability and structure of S-doped monolayer and few-layer graphene}
\label{Sec:en_S_doping}
\begin{figure}[t]
    \includegraphics[width=0.95\linewidth]{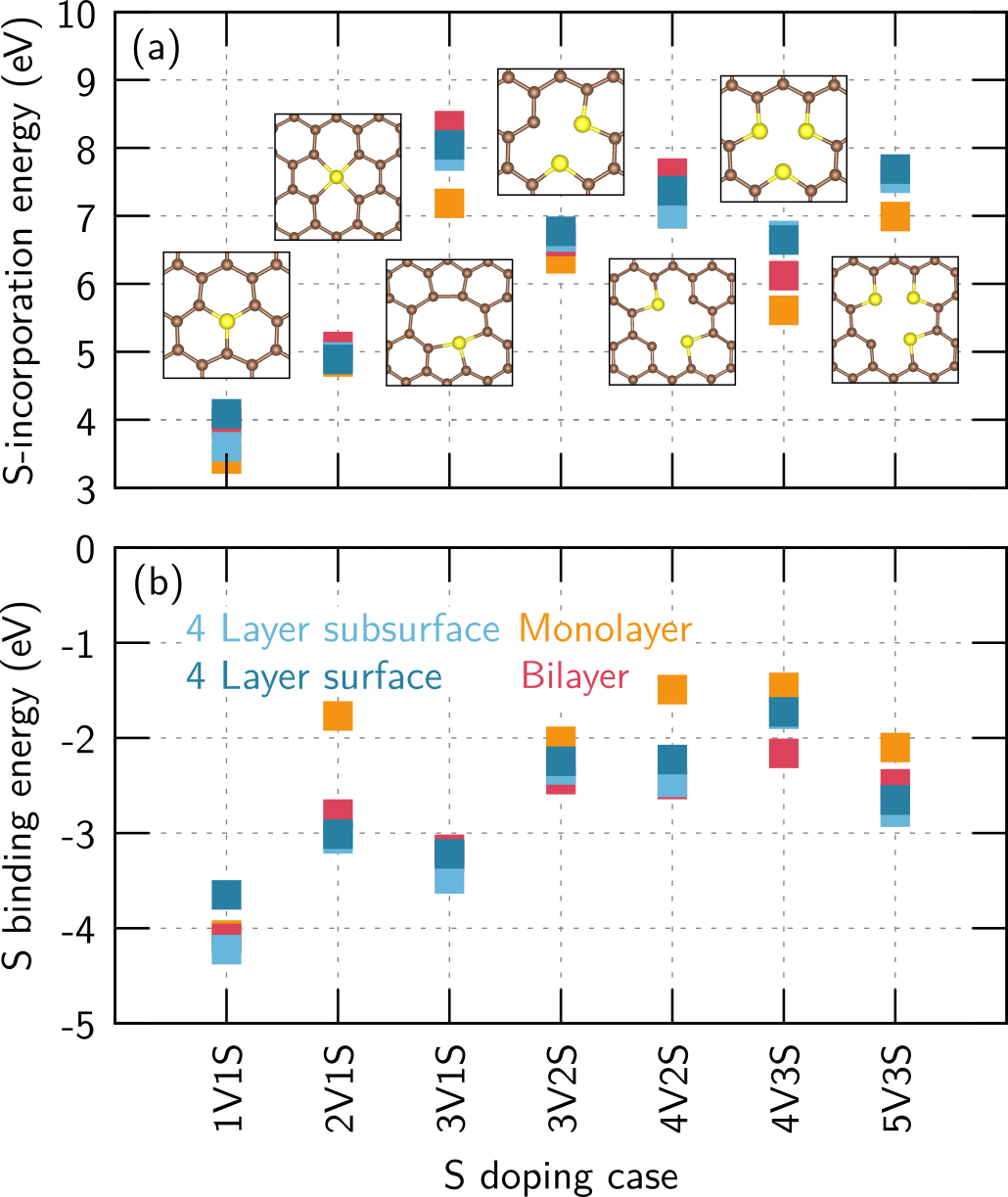}
    \caption{(a) S-incorporation energy of the considered doping configurations. The insets show the top view of the relaxed S-doped configuration. (b) Binding energies of S-dopants at the vacancy shown in Fig.~\ref{fig:structure_vacancy}a. Orange, red, light blue and dark blue represent the monolayer, bilayer, subsurface and surface doped 4L-FLG cases, respectively. }
    \label{fig:stability}
\end{figure}
We now turn to the effect of S-doping and focus on the structure and stability of S-doped graphene. 
Insight in the latter is obtained from the S-incorporation energy which renders the energy cost associated with the doping of pristine graphene (cf. Eq.~\ref{eq:incorporation}), whereas the binding energy represents the energy to bind the dopant to a preexisting vacancy (cf. Eq.~\ref{eq:binding}). 
Both quantities for each case in monolayer, bilayer and four-layer graphene are displayed in Fig.~\ref{fig:stability}a and b, respectively, as well as in  Table~\ref{tab:form_en}. 
In 4L-FLG we explore explicitly the layer-selective incorporation in either the surface or subsurface layer and its implications. 
Owing to the higher formation energies of vacancies in the multilayer, the S-incorporation energy in FLG is higher for all studied configurations (cf. Fig.~\ref{fig:stability}a and Table~\ref{tab:form_en}). 
On the other hand, all binding energies are negative, signifying that it is energetically favorable to bind S in an existing vacancy.  
Moreover, the binding energy  of all doping configurations is enhanced in multilayer systems compared to monolayer graphene, which underscores the role  of the surrounding graphene layers in FLG.

\begin{table}                                                                   
\centering                                                                       
    \begin{ruledtabular}                                                         
    \caption{Out-of-plane displacement of the S-dopant with respect to the graphene plane in \AA{}. Negative/positive values relate to an inward/outward displacement of the S-dopant, respectively. }
        \begin{tabular}{lcccc}                                               
        Case & \multicolumn{4}{c}{  Vertical relaxation of S [\AA] } \\
        & 1L & 2L & 4L-subsurface & 4L-surface   \\
        \hline
        1V1S  &$1.15$ &$0.9$ &$0.89$ &$0.78$ \\
        2V1S  &$0$    &$0$    &$0$             &$0$          \\
        3V1S  &$0.74$  &$0.77$  &$0.55$           &$-0.42$          \\
        3V2S  &$1.17$  &$0.85$  &$0.62$           &$-0.69$          \\
        4V2S  &$0.44$  &$0.37$  &$0.29$           &$-0.11$          \\
        4V3S  &$1.47$  &$1.41$  &$1.06$           &$-1.30$          \\
        5V3S  &$1.32$  &$0.95$  &$0.88$           &$-1.45$          \\
        \end{tabular}
        \label{tab:displacement}
    \end{ruledtabular}
\end{table}

\begin{table*}                                                                   
\centering                                                                       
    \begin{ruledtabular}                                                         
    \caption{S-incorporation energy and binding energy of the S-dopant onto a vacancy in eV for all considered cases. The lowest energy and strongest binding configuration is marked in bold numbers.}
        \begin{tabular}{lcccccccc}                                               
        Case & \multicolumn{4}{c}{  S incorporation energy (eV) } &\multicolumn{4}{c}{ S binding energy (eV) } \\
        & 1L & 2L & 4L-subsurface & 4L-surface & 1L & 2L & 4L-subsurface & 4L-surface  \\
        \hline
        1V1S  &$\mathbf{3.42}$ &$\mathbf{3.96}$ &$\mathbf{3.60}$ &$\mathbf{4.09}$ &$\mathbf{-4.07}$ &$\mathbf{-4.11}$ &$\mathbf{-4.22}$ &$\mathbf{-3.65}$ \\
        2V1S  &$4.85$          &$5.08$          &$4.92$          &$4.89$          &$-1.77$          &$-2.80$          &$-3.07$          &$-3.01$          \\
        3V1S  &$7.18$          &$8.33$          &$7.88$          &$8.04$          &$-3.25$          &$-3.17$          &$-3.48$          &$-3.22$          \\
        3V2S  &$6.37$          &$6.62$          &$6.69$          &$6.77$          &$-2.03$          &$-2.43$          &$-2.34$          &$-2.25$          \\
        4V2S  &$7.03$          &$7.63$          &$7.03$          &$7.37$          &$-1.49$          &$-2.49$          &$-2.47$          &$-2.23$          \\
        4V3S  &$5.61$          &$6.12$          &$6.71$          &$6.64$          &$-1.47$          &$-2.16$          &$-1.75$          &$-1.73$          \\
        5V3S  &$6.98$          &$7.61$          &$7.55$          &$7.69$          &$-2.10$          &$-2.48$          &$-2.78$          &$-2.65$          \\
        \end{tabular}
        \label{tab:form_en}
    \end{ruledtabular}
\end{table*}

\begin{figure}[t]
    \includegraphics[width=1\linewidth]{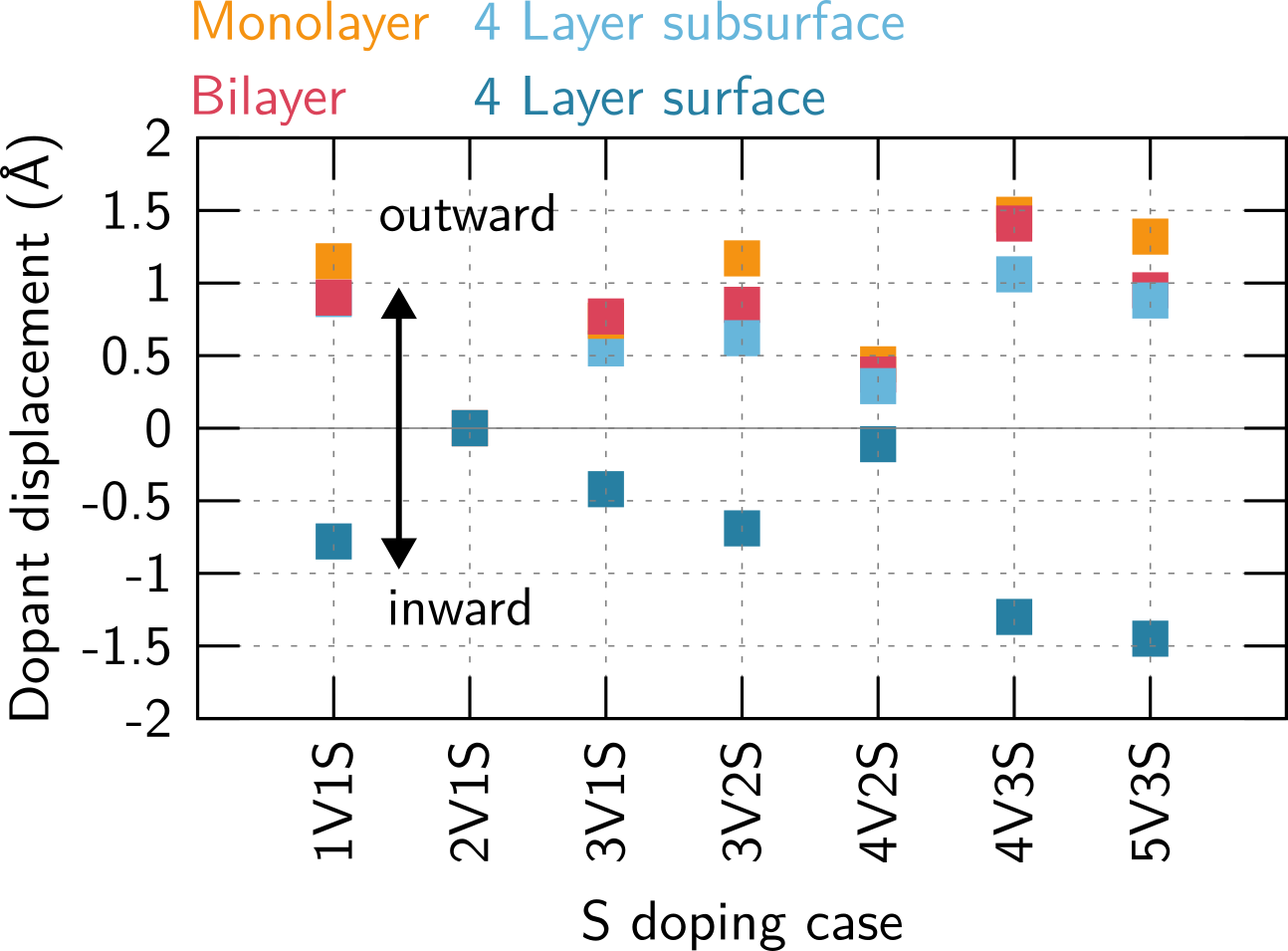}
    \caption{Out-of-plane displacement of the S-dopant with respect to the graphene plane. The colors orange, red, light blue and dark blue represent the S-dopant out-of-plane displacement in a monolayer, bilayer, subsurface and surface doped 4L-FLG, respectively. Negative values and positive values relate to an inward and outward displacement of the S-dopant, respectively.  }
    \label{fig:structure}
\end{figure}

According to the S-incorporation energy, graphitic 1V1S is the most stable configuration (3.42~eV). 
This also holds for few-layer graphene with an S-incorporation energy of 3.96~eV, 3.60~eV and 4.09~eV in the bilayer, subsurface and surface doped 4L-FLG, respectively. 
This finding is corroborated by the strong binding of -4.07~eV, -4.11~eV, -4.22~eV and -3.65~eV of the S-dopant to the vacancy in monolayer, bilayer, subsurface and surface doped 4L-FLG, respectively (cf. Fig.~\ref{fig:stability}b and Table~\ref{tab:form_en}).
Most importantly, the S-dopant exhibits a strong out-of-plane displacement of 1.15~\AA{}, 0.9~\AA{}, 0.89~\AA{} and -0.78~\AA{} with respect to the graphene plane in the monolayer, bilayer, subsurface and surface-doped 4L-FLG (cf. Table~\ref{tab:displacement} and Fig.~\ref{fig:structure}). 
This leads to a C-S bond-length of 1.74~\AA{} in monolayer and bilayer graphene, and 1.72~\AA{} in subsurface and surface doped 4L-FLG.

The second most stable configuration is thiophenic 2V1S with an S-incorporation energy of 4.85~eV, 5.08~eV, 4.89~eV and 4.92~eV, in monolayer, bilayer, subsurface and surface doped 4L-FLG (cf. Table~\ref{tab:form_en}). 
The sulfur forms a unique bond configuration with four bonds to neighboring C-atoms resulting in a structure with two 5-rings and two 6-rings. In contrast to graphitic incorporation, this configuration is planar without a vertical relaxation of the S-dopant. 
The C-S bond length of $1.87$~\AA\ in monolayer, bilayer and 4L-FLG is characteristic of a C-S single bond in thiol~\cite{TRINAJSTIC_C-S_bonds}. 
The S-binding energy in 2V1S is $-1.77$ eV, $-2.80$ eV, $-3.07$ eV and $-3.65$ eV for monolayer, bilayer, subsurface and surface doped 4L-FLG, respectively, indicating a significant enhancement of the binding energy in the multilayer  system. 

Thiopyranic 4V3S, comprising three S-dopants in a four site vacancy, is the third most stable configuration with a S-incorporation energy of $5.61$~eV, $6.12$~eV, $6.71$ eV and $6.64$~eV in the monolayer, bilayer, surface and subsurface layer of 4L-FLG (cf. Table~\ref{tab:form_en}). Alternatively, the binding energy of $-1.47$ eV (1L), $-2.16$ eV (2L), $-1.75$ eV (subsurface 4L-FLG) and $-1.73$ eV (surface doped 4L-FLG), cf. Fig.~\ref{fig:stability}a and Table~\ref{tab:form_en}) indicates an overall weaker binding of the S-dopants to the vacancy compared to 1V1S and 2V1S. A considerable S-dopant out-of-plane displacement of 1.47~\AA{}, 1.41~\AA{}, 1.06~\AA{} and -~1.30~\AA{} is found in the monolayer, bilayer and subsurface and surface doped 4L-FLG, respectively (cf. Fig.~\ref{fig:structure} and Table~\ref{tab:displacement}). This goes hand in hand with an enhanced C-S bond-length of $1.73$ \AA, $1.72$ \AA, $1.69$ \AA{} and $1.70$ \AA{} in the monolayer, bilayer, subsurface and surface doped 4L-FLG, respectively,  signaling a bond configuration between $sp^2$ and $sp^3$.

While previous studies have often considered planar geometries~\cite{tucek_sulfur_2016, Wang-S-doped_graphene_Li:2019}, our results indicate significant vertical relaxation of S up to $1.47$~\AA{} (cf. Fig.~\ref{fig:structure}a and Table~\ref{tab:displacement}). 
Specifically, in the bilayer and subsurface doped 4L-FLG the S-displacement is in the outward direction, while in the surface doped 4L-FLG an inward relaxation of 1.45~\AA\ occurs.
The vertical displacement goes hand in hand with a substantial gain of energy, e.g., for graphitic 1V1S doping in the monolayer the incorporation energy is lowered by 2.09 eV. 
The energy gain due to the S-dopant out-of-plane relaxation ranges between 0.33 and 2.09~eV, as described in Appendix~\ref{app:outofplane}. 

Last but not least, we explored also the binding energy of an S-atom at the pristine graphene surface, as such configurations as well as intercalation 
are of interest for application as a cathode material in batteries~\cite{Pope_Li_S_batteries_2015, Yang2021_graphene_battery_sulfur}. 
We find that the most stable adsorption configuration is an S-atom at a bridge position between two C-atoms. 
Overall, the binding energies are positive (since in Eq.~\ref{eq:binding} an S$_8$-ring is used as the reference system): $2.25$ eV, $1.50$ eV and $1.38$ eV for adsorption at the surface of the monolayer, bilayer and 4L-FLG, respectively, indicating that this configuration is energetically less favorable than adsorption in vacancies. 
Additionally, the binding energy of an S-atom intercalated between the sheets of 4L-FLG is $2.08$ eV.

\begin{figure*}[t]
    \includegraphics[width=1\linewidth]{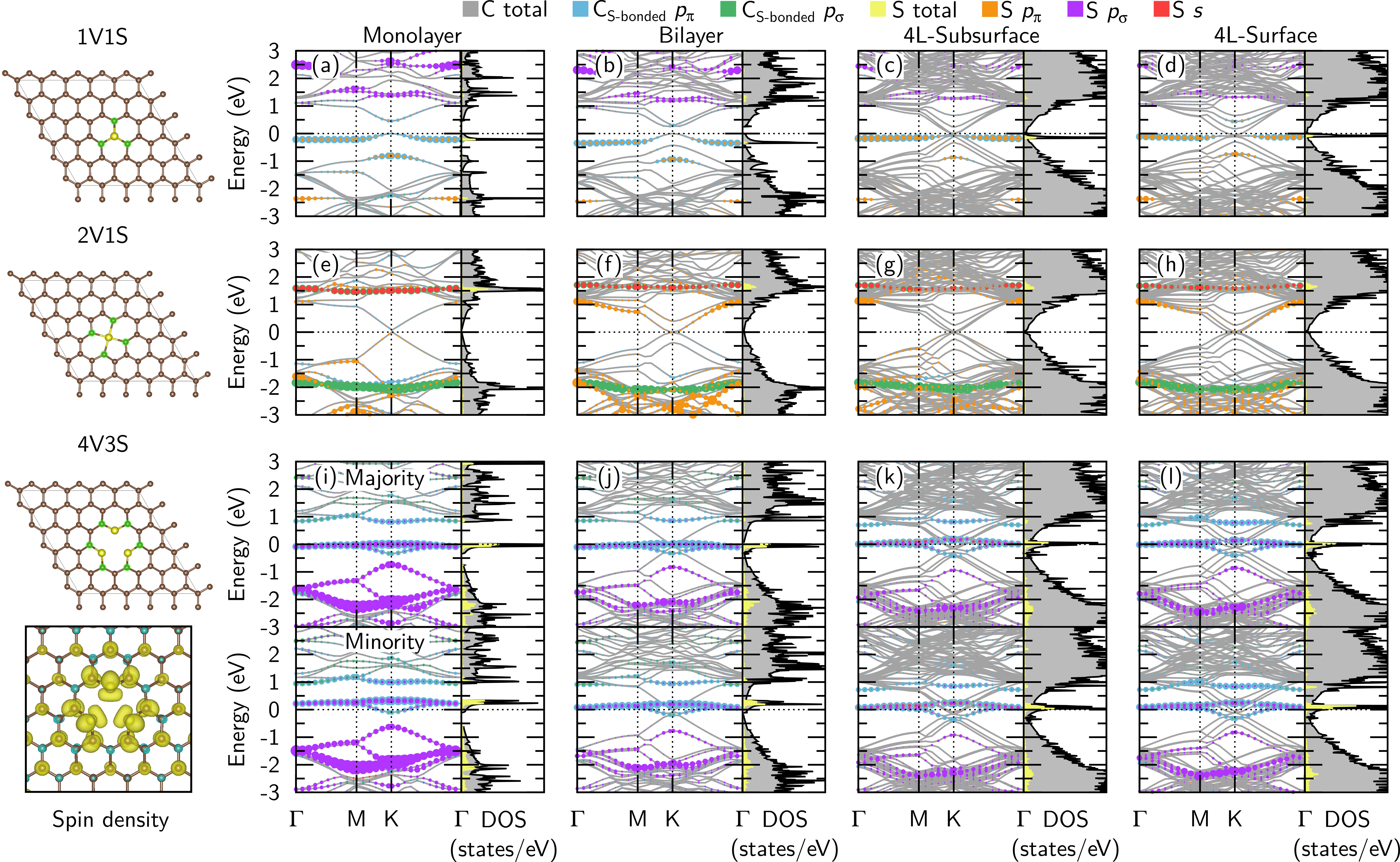}
    \caption{
    Element- and orbital-resolved band structure of the doped layer and total and S-contribution to the density of states of  the (a,b,c,d) 1V1S, (e,f,g,h) 2V1S  and  (i,j,k,l) 4V3S  configurations in monolayer, bilayer, subsurface and surface doped 4L-FLG structures. 
    In each row a top view of the presented configuration is shown on the left.
    C-atoms neighboring the S-dopant are showed in green.
    The majority and minority bands of 4V3S are displayed in separate panels. 
    The spin density of the 4V3S case is shown with an isosurface of $2.5\cdot10^{-3}$~$\text{e}/\text{\AA}^3$. The yellow and cyan color depict the majority and minority spin-density, respectively. 
    C $p_\pi$ and C $p_\sigma$ of the C-neighbors of the dopant, S $p_\pi$, S $p_\sigma$ and S $s$ contribution to the band structure are shown in blue, green, orange, purple and red, respectively. The contribution of all C-atoms in the system is shown in light gray. The S-state projection in the density of states is shown in yellow. 
    }
    \label{fig:bands}
\end{figure*}

\section{Electronic structure of S-doped monolayer and few-layer graphene}
\label{Sec:Bands}
Having identified  1V1S, 2V1S and 4V3S as the most stable doping configurations, we focus here on their electronic properties.
The element-, orbital- and layer-resolved band structures of S incorporated in a monolayer, bilayer and in the surface or subsurface layer of 4L-FLG are shown in Fig.~\ref{fig:bands}. 
Additionally, the unfolded band structures are presented in Fig.~\ref{fig:bands-unfolded}.

The band structure of the most stable graphitic 1V1S doping configuration in the monolayer, shown in Fig.~\ref{fig:bands}a, can be rationalized by comparing to the one of 1V (Fig.~\ref{fig:structure_vacancy}d): 
The sublattice imbalance leads to a similar band gap of 0.4~eV between hyperbolic conduction and valence bands.
In contrast to 1V, the valence band is completely shifted below the Fermi level and the flat region away from the K-point towards $\Gamma$ and M shows now a pronounced $p_\pi$ contribution of S and the neighboring C.  
This in-gap defect band leads to a peak in the total DOS just below the Fermi level (-0.21~eV).  
A further band with reduced dispersion is found below -0.7~eV. 
Its maximum at K exhibits a mixed S and C $p_\pi$ character, whereas impurity bands are formed with S $p_\sigma$ character at 1.5 and 2.5~eV above the Fermi level.
We also observe a profound effect of the vertical relaxation on the band structure: while the 2.09 eV less stable planar 1V1S configuration can be understood in terms of $n$-type doping of the Dirac cone, a strong rearrangement of the band structure emerges due to the vertical displacement of the dopant (cf. Fig.~\ref{fig:structure}) characteristic of a modification from $sp^2$ towards $sp^3$-like hybridization. 
A splitting of the Dirac cone is also found in graphitic N- or B-doping, however, it is accompanied by a shift of the valence band above the Fermi-level ($p$-doping for B) or of the conduction band below the Fermi-level ($n$-doping for N)~\cite{ferrighi_boron-doped_2015}. 
In 1V1S the additional carriers introduced by the S-doping saturate the dangling bonds of 1V and not only quench the $p$-doping, but also the spin polarization of 1V, consistent with Ref.~\cite{zhu_magnetic_2016}. 
The lack of spin-polarization in 1V1S may also be related to the lower S-concentration of 2\% considered here, in contrast to earlier work that reported spin-polarization for graphitic doping with S-concentration in the range of 4\%-6\%~\cite{tucek_sulfur_2016}.

1V1S incorporated in the graphene bilayer (cf. Fig.~\ref{fig:bands}b) shows a similar band structure to the monolayer case. The band gap persists, albeit it is reduced to 0.25~eV, separating two hyperbolic conduction bands with slightly smaller shift away from K than 1V (Fig.~\ref{fig:structure_vacancy}e) and a valence band with a maximum at K, whose flat region away from K lies lower than in the monolayer case. 
On the other hand, in the 1V1S 4L-FLG case both for subsurface (cf. Fig.~\ref{fig:bands}c) and surface doping (cf. Fig.~\ref{fig:bands}d) this band becomes even flatter with a stronger S-character and is shifted closer to the Fermi level, leading to a high peak in the DOS at $E_{\rm F}$. 4L-FLG contributes four pairs of hyperbolic bands, that are more dispersive for subsurface doping, with a more graphene-like character for the undoped surface layer. One (two) pairs of the hyperbolic bands are touching at $E_{\rm F}$ for surface (subsurface) doping and the remaining show increasing splitting and reduced dispersion. 
While the band structure of the doped layer shows similar features both in the monolayer and multilayer, the latter is a superposition of the band structure of the undoped layers and  the doped layer. Overall, the interlayer interaction and the reduced doping concentration leads to a band gap reduction and finally closing for 4L-FLG. 
Thus, our results demonstrate that both the band gap and the position and dispersion of the impurity band near the Fermi-level is sensitive to and can be modified by the number of layers in the FLG. 

Strikingly, in the thiophenic 2V1S doping configuration in the monolayer (cf. Fig.~\ref{fig:bands}e) the Dirac-crossing at the Fermi level and its linear dispersion are largely preserved. 
This can be rationalized by considering that in the 2V1S configuration two C-atoms with 4 electrons each are replaced by one S-dopant with 6 electrons. 
Hence, the doped graphene is two-electron deficient. 
However, due to the structural rearrangement no $p$-doping is found since S renders 4 electrons to form four covalent bonds with its C-neighbors, preserving the Dirac cone of the unperturbed part of the lattice. 
An occupied flat band with C $p_\sigma$ character of the direct neighbors of the S-dopant appears $-2$ eV below the Fermi-level hybridizing with more dispersive S $p_\pi$ bands, and an empty impurity flat band of S $s$ character lies 1.5 eV above the Fermi-level. 

The main difference in the electronic structure when the 2V1S is embedded in a bilayer or the subsurface/surface layer of 4L-FLG (cf. Fig.~\ref{fig:bands}f-h) is the modification of the Dirac cone to hyperbolic valence and conduction pairs of bands, two in the bilayer and four in 4L-FLG, with one pair (bilayer)/two pairs (4L-FLG) touching at $E_{\rm F}$,  reminiscent of the ones of undoped few-layer graphene (cf. Fig.~\ref{fig:bands_pristine_FLG}, as well as Ref.~\cite{mak_evolution_2010}). Besides the two impurity bands, the occupied band at -2~eV contributed by the nearest C-neighbors to S and an empty one of S $s$ character at 1.7~eV, observed already for the monolayer, states of S $p_\pi$ character appear below -2~eV as well as a band at 1.1~eV. 
Overall, thiophenic 2V1S emerges as a promising way to incorporate S into (multilayer) graphene without lifting the Dirac cone.

Next, we turn to the thiopyranic 4V3S configuration (cf. Fig.~\ref{fig:bands}i). Most notably, the broken sublattice symmetry leads to a spin-polarization, where the magnetic solution with a total magnetic moment of 0.99~$\mu_\text{B}$ is favored by 49~meV per simulation cell. Additionally, a split Dirac cone emerges with a more dispersive valence band with maximum at -0.6~eV at K with predominantly S $p_\sigma$ character and a relatively flat conduction band of $p_\pi$ origin of the C neighbors at 0.7~eV (majority) and 0.5~eV (minority channel). 
A prominent feature is the impurity flat band of mixed S $p_\sigma$ and neighboring C $p_\pi$ character lying at the Fermi level in the majority spin channel and slightly higher for the minority spin, leading to a peak in the DOS close to the Fermi level. 
A slightly more dispersive conduction band of C $p_\pi$ nature with a minimum at K lies 0.3~eV below $E_{\rm F}$ in the majority spin channel and touches the latter in the minority spin channel.  
The $n$-type doping, which is more pronounced in the majority spin channel, can be rationalized by the fact that the vacancy creates $4$ holes, whereas the three S-dopants donate $2$ electrons each, leading to a surplus of $2$ electrons. 

The magnetic moment of 0.99~\mub\ stems from the exchange splitting of the flat and slightly more dispersive bands around $E_{\rm F}$. 
The largest contribution to the spin-density shown on the left in Fig.~\ref{fig:bands} is localized at the S-sites and associated with a  magnetic moment of $0.08\ \mu_\text{B}$ per S. 
The contribution of the nearest and further neighbors on the same sublattice is weaker and decreasing with increasing distance. 
Additionally, the sites on the other sublattice render a small but negative contribution. 
Overall, the spin density is much more delocalized than the one in the vacancy configurations (1V, 3V, 5V) discussed in Section~\ref{Sec:Vac} (cf. Fig.~\ref{fig:structure_vacancy}a). 

The general characteristics of the band structure of the 4V3S monolayer (cf. Fig.~\ref{fig:bands}j-l) are preserved for the multilayer case with less  S $p_\sigma$ character in the valence band.
The main difference in the 4V3S bilayer configuration is that a Dirac cone from the intact graphene layer is superimposed on the band structure of the doped layer (cf. Fig.~\ref{fig:bands}j). 
Notably, this  demonstrates that it is possible to couple the linear dispersion of the Dirac cone in the undoped layer with the impurity band features of the doped layer. 
The 4V3S 4L-FLG cases (cf. Fig.~\ref{fig:bands}k, l) show additionally three pairs of hyperbolic valence and conduction bands. One pair is touching below $E_{\rm F}$, indicating $n$-type doping and the rest are split apart, the splitting being larger for surface doping. 
An intriguing feature is that the $n$-type doping is not confined to the doped layer but extends to the undoped layers, emphasizing the interaction and charge transfer between the doped and undoped layers. 
Thus, the 4V3S configuration in 4L-FLG allows one to combine the impurity flat band feature of the doped monolayer with the hyperbolic bands of the multilayer, achieving an overall $n$-type doping.  
While the total magnetic moment of the 4V3S bilayer case (0.95 $\mu_\text{B}$) is similar to the one of the monolayer, it is significantly reduced in case in 4L-FLG to  0.39~$\muB$ and 0.27~$\muB$ for surface and subsurface doping, respectively, consistent thus smaller exchange splitting between the impurity bands around the Fermi level in 4L-FLG. 

\section{Summary}

We present a systematic investigation of the effect of S-doping on the structure, stability and electronic properties of mono-, bi-, and four-layer graphene using density functional theory including van der Waals corrections. 
As a starting point, we considered the properties of vacancies in single and few-layer graphene. 
We found that the formation energy per site is reduced from 7.49~eV (1V) to 3.32 eV (2V) and,  finally, to 2.65 eV (5V), making the formation of larger vacancies feasible. 
Moreover, spin polarization of 1.0-1.5~\mub\ was found in 1V, 3V and 5V localized at the sites with unsaturated bonds and their neighbors, in contrast to 2V and 4V due to saturated dangling bonds that also lead to a higher stability.

A variety of doping configurations was considered including  graphitic, thiophenic and thiopyranic. 
We showcase here the most stable ones, 1V1S, 2V1S and 4V3S, which exhibit distinct features in the band structure. 
Except for 2V1S, we find a strong out-of-plane relaxation of the dopant that impacts the electronic properties. 
For example, while a strong $n$-doping doping of the split Dirac cone is found for the metastable planar 1V1S configuration, the band structure of the 2.09~eV more stable non-planar graphitic 1V1S  features a split Dirac cone with a band gap of 0.4~eV, in conjunction with an impurity flat band of S- and C-$p_\pi$ character at the Fermi-level. The $p$-type doping and the spin-polarization of the 1V vacancy are quenched due to the additional charge carriers introduced by S, that saturate the dangling bond. 

A similar band structure with a split Dirac cone and additional nearly  flat bands around the Fermi level with substantial S $p_\sigma$ and neighboring C $p_\pi$ contribution is found in 4V3S. Most importantly, the low-dispersion bands at \ef\ show  $n$-type doping and a notable exchange splitting, leading to a magnetic moment of 0.99~\mub\ and a spin-density that is much more delocalized than the one in vacancies. 

Finally, thiophenic 2V1S represents a configuration where the Dirac cone is sustained. Additional empty and occupied impurity flat bands emerge, contributed by S and the neighboring C, respectively. Here the S-dopant renders two $p_\pi$ electrons to graphene, whereas the remaining four electrons form a tetravalent bond with four neighboring C. 

A central aspect of our study is to explore how the  properties of the defective monolayer are altered in the multilayer. 
For 1V instead of the split Dirac cone of the monolayer,  two pairs of hyperbolic bands with VBM and CBM  shifted away from K emerge in the bilayer, whereas 4L-FLG features four pairs of hyperbolic bands, one touching at \ef. Moreover, the top valence band becomes flatter and the $p$-type doping is reduced with increasing thickness. 
Similarly, a reduction of the band gap with a smaller shift of the conduction band minima away from K is observed in the 1V1S bilayer and a closing of the band gap in 4L-FLG. Concomitantly, the dispersion of the impurity-derived band at the Fermi level is further reduced in 4L-FLG. 
Remarkably, in the 4V3S bilayer a Dirac cone with linear dispersion originating from the intact graphene layer is overlaid  on the  band structure  of the the doped layer. 
For 4V3S in 4L-FLG, the undoped layers render hyperbolic touching bands that show  $n$-doping. This is coupled with a reduced exchange splitting of the impurity bands that decreases the magnetic moment to 0.39~\mub\ and 0.27~\mub\ in surface and subsurface doping, respectively. 
Finally, in 2V1S the Dirac cone is altered into two/four pairs of hyperbolic bands pairs, one/two touching at \ef, reminiscent of undoped few-layer graphene~\cite{mak_evolution_2010}, while the splitting of the occupied and empty impurity flat bands is reduced due to  screening.

Overall, our results demonstrate that a wide range of features can be realized in the band structure of S-doped graphene depending on the configuration, ranging from sustained Dirac cone (2V1S) to split Dirac cone in conjunction with the emergence of impurity flat bands at the Fermi level in graphitic 1V1S that are also exchange split in 4V3S.  
Beyond twisted bilayer and rhombohedral graphene~\cite{Munoz_rhombohedral_2013, tian_evidence_2023}, impurity flat bands induced e.g. by vacancies in graphene  have been recently proposed to host strong correlations and even superconductivity~\cite{vanpoppelen2026disorderinducedsuperconductivitygraphene, Marsal_enhanced_quantum_metric_2024}.
In this context the broad tunability of the electronic properties of S-doped graphene and the emergence of impurity derived flat bands allows to design the band structure by selective incorporation of S and has potential for a wide array of applications in catalysis, energy storage and spintronics.

\begin{acknowledgments}
We acknowledge funding by the Deutsche Forschungsgemeinschaft (DFG, German Research Foundation) - IRTG 2803 - 461605777
and the National Science Foundation, Grant No.~NSF-DMR-2118718.
Computing time was granted by the Center for Computational Sciences and Simulation of the University of Duisburg-Essen
(DFG Grants No.~INST 20876/209-1 FUGG and No.~INST 20876/243-1 FUGG).
\end{acknowledgments}

\appendix

\section{Undoped few-layer graphene band structures}
\label{app:pristine-FLG}

\begin{figure}[h]
    \includegraphics[width=0.8\linewidth]{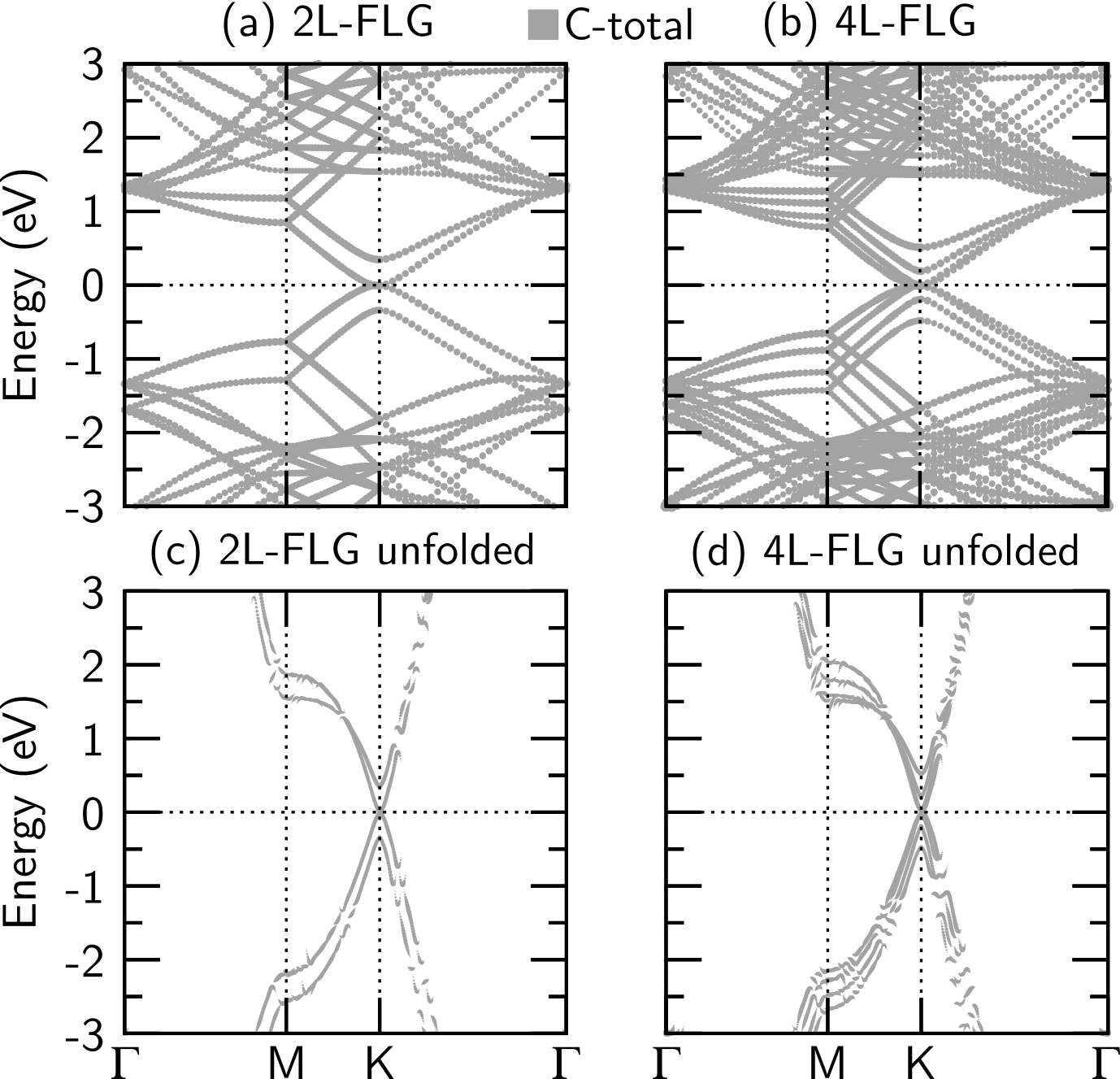}
    \caption{ Band structure of the undoped and pristine (a) bilayer and (b) 4L-FLG in a $5\times5$ supercell. Additionally, the band structure unfolded to the primitive unit-cell of (c) bilayer and (d) 4L-FLG is shown. The gray color signifies the total contribution of C to the band structure. 
    }
    \label{fig:bands_pristine_FLG}
\end{figure} 

\begin{figure*}[!htbp]
    \includegraphics[width=1\linewidth]{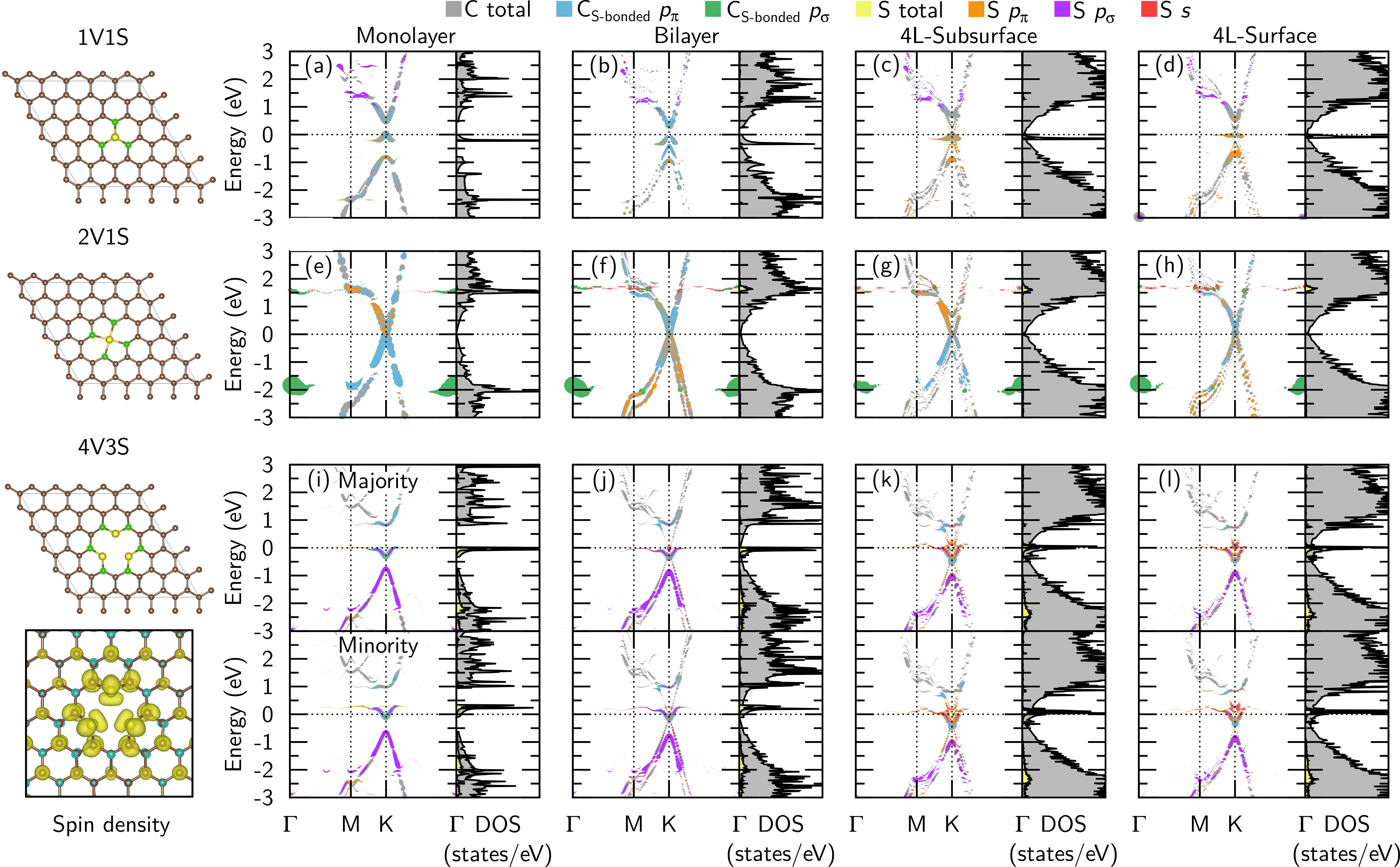}
    \caption{
    Element-, orbital- and layer-resolved band structure of the doped layer and total and S-contribution to the density of states of 
    the (a,b,c,d) 1V1S,  (e,f,g,h) 2V1S  and  (i,j,k,l) 4V3S  configurations in monolayer, bilayer, subsurface and surface doped 4L-FLG structures. The projection was scaled by a factor of $20$ and $6$ for S and neighboring C, respectively, up to make the contribution of the orbitals visible. 
    In each row a top view of the presented configuration is shown on the left.
    C-atoms neighboring S-dopant are showed in green.
    The band structures of the $5\times5$ supercell were unfolded to the $1\times1$ primitive cell. The majority and minority bands of 4V3S are shown in separate panels. 
    The spin density of the 4V3S case is shown with an isosurface of $2.5\cdot10^{-3}$~$\text{e}/\text{\AA}^3$. The yellow and cyan color depict the majority and minority spin-density, respectively. 
    C contributions of the doped layer are shown in gray. 
    C $p_\pi$ and C $p_\sigma$ of the C-atoms neighboring the dopant, S $p_\pi$, S $p_\sigma$ and S $s$ contribution to the band structure is shown in blue, green, orange, purple and red, respectively. 
    The S-state projection in the density of states is shown in yellow. 
    }
    \label{fig:bands-unfolded}
\end{figure*} 
We elaborate briefly on the main features of the undoped graphene bilayer and 4L-FLG band structure shown in ~\ref{fig:bands_pristine_FLG}a and b, respectively. 
The interlayer coupling and the altered vertical $p_\pi$ bonds lead to a modification of the linear dispersion of the Dirac cone into two pairs of valence and conduction hyperbolic bands in the bilayer (cf. Fig.~\ref{fig:bands_pristine_FLG}a), one pair touching at the Fermi-level and the remaining split apart~\cite{mak_evolution_2010}. 
Similarly, we find four hyperbolic pairs in 4L-FLG, two pairs touching at the Fermi-level and the remaining two pairs split apart.  
The splitting varies with the layer number $N$ and position in the layer stacking sequence, with a smaller energy splitting in near-surface layers.

\section{Unfolded Band Structures}

\begin{figure}[t]
    \includegraphics[width=1\linewidth]{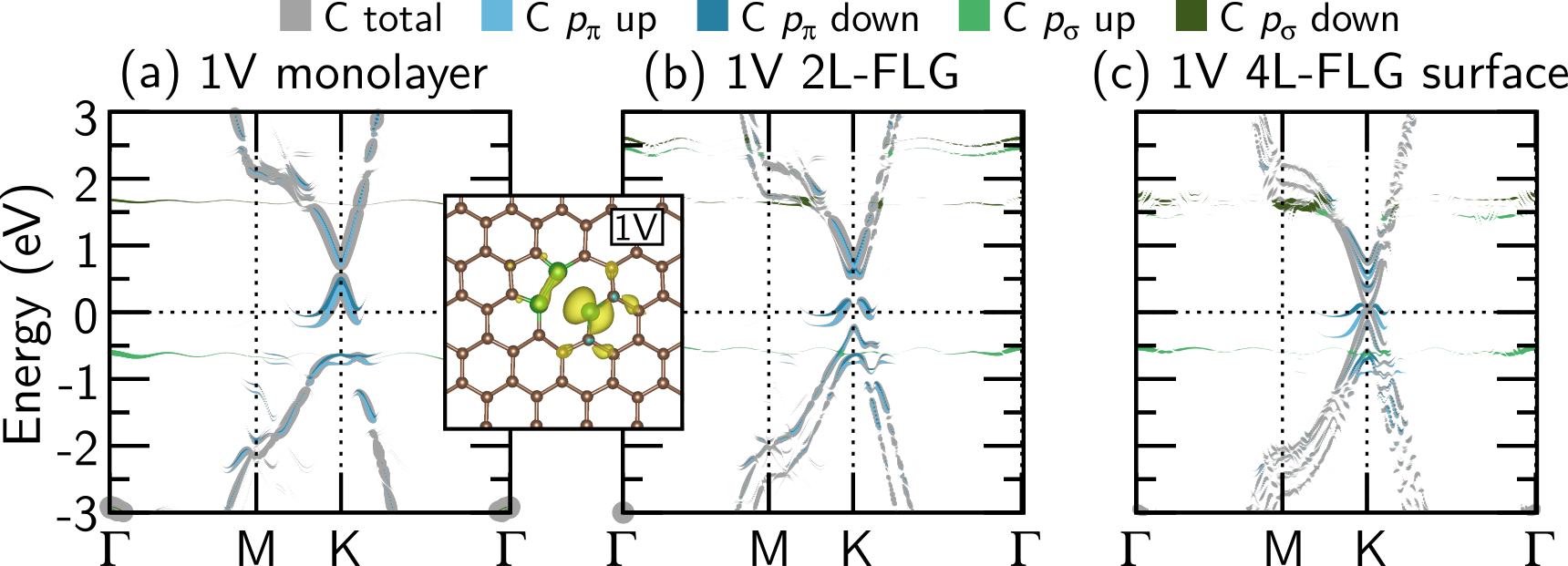}
    \caption{
    Element-, orbital- and spin-resolved band structure of the 1V single vacancy in (a) the monolayer, (b) the bilayer and (c) the surface layer of 4L-FLG, unfolded to the $1\times1$ primitive cell. 
    In an inset the top view of the monovacancy is shown. Additionally, C-atoms neighboring the vacancy are shown in green.   
    C contributions of the defective layer are shown in gray. 
    The majority/minority C $p_{\sigma}$ and C $p_{\pi}$ associated with the C-atoms marked in green are shown in green/dark green and blue/dark blue, respectively. 
      }
    \label{fig:vacancy-unfolded}
\end{figure}

Complementary to the 1V band structure in the monolayer and in few-layer graphene displayed in Fig.~\ref{fig:structure_vacancy}, in Fig.~\ref{fig:vacancy-unfolded} we  show the layer-, element- and orbital-resolved band structures unfolded  to the Brillouin zone of the primitive cell. Likewise, Fig.~\ref{fig:bands-unfolded} renders the corresponding unfolded band structures of 1V1S graphitic, 2V1S thiophenic and 4V3S.
Note that in Fig.~\ref{fig:bands-unfolded} only the contribution of the layer containing the dopant is shown.

\section{Effects of S-out-of-plane relaxations }
\label{app:outofplane}
\begin{figure}[h]
    \includegraphics[width=0.8\linewidth]{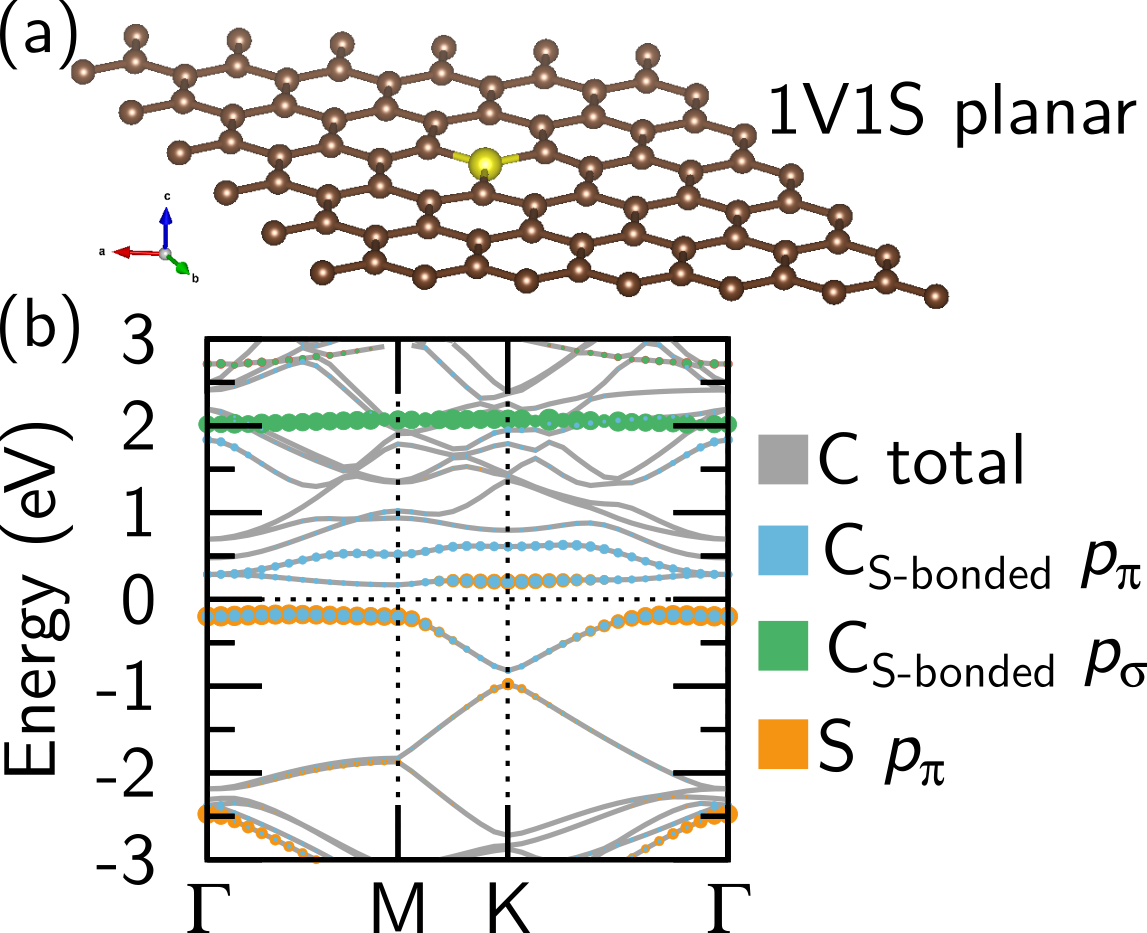}
    \caption{ (a) Tilted view of the planar 1V1S structure. (b) Element- and orbital-resolved band structure of the planar monolayer 1V1S. 
    C $p_\pi$ and C $p_\sigma$ of the C-neighbors of the dopant, and S $p_\pi$ contribution to the band structure are shown in blue, green, orange, respectively. The contribution of all C-atoms in the system is shown in light gray.
    }
    \label{fig:folded_1V1S_planar}
\end{figure}

Next, we show the importance of considering out-of-plane relaxation in the S-doped 1V1S, 3V1S, 3V2S, 4V2S, 4V3S and 5V3S (cf. Fig.~\ref{fig:stability}) to obtain the correct ground state. As shown in Fig.~\ref{fig:meta-stable}l, m, o-r, the S-doped configuration with a planar geometry are associated with an increase of 2.09~eV, 1.60~eV, 0.53~eV, 0.33~eV, 1.65~eV and 0.99~eV in energy for 1V1S, 3V1S, 3V2S, 4V2S, 4V3S and 5V3S, respectively, compared to the cases with out-of-plane relaxation (cf. Fig.~\ref{fig:stability} and Table~\ref{tab:form_en}). The 2V1S configuration with out-of-plane relaxation of the S-dopant is shown in Fig.~\ref{fig:meta-stable}n and is associated with an increase of 1.48~eV in total energy.

The band structure of the planar 1V1S configuration (cf. Fig.~\ref{fig:folded_1V1S_planar}b) differs significantly from the one of the 2.09 eV more stable non-planar 1V1S (cf. Fig.~\ref{fig:bands}a). Due to the electron doping the split Dirac cone is shifted completely below the Fermi level. A band gap opens between the now occupied conduction band that flattens out towards $\Gamma$ which has a strong contribution of S $p_\pi$ states and neighboring C and an empty band with more pronounced C-character of the neighboring sites. Furthermore, a defect level with $p_\sigma$ character of the neighboring C is found at 2 eV above the Fermi-level. A significant rearrangement is observed for the non-planar 1V1S configuration (cf. Fig.~\ref{fig:bands}a), as described in the main text. 

\section{Metastable Configurations}
\label{app:metastable}

Reconstructions of the C-network may lead to multiple distinct configurations of defects in graphene~\cite{Banhart:2011, Kotakoski_large_vacancies_2014, Kotakoski_point_defects_2011}. Here we present several metastable defect structures in the monolayer and their relative energies to the lowest energy configuration, described in the main text. 

\begin{figure}[h]
    \includegraphics[width=1\linewidth]{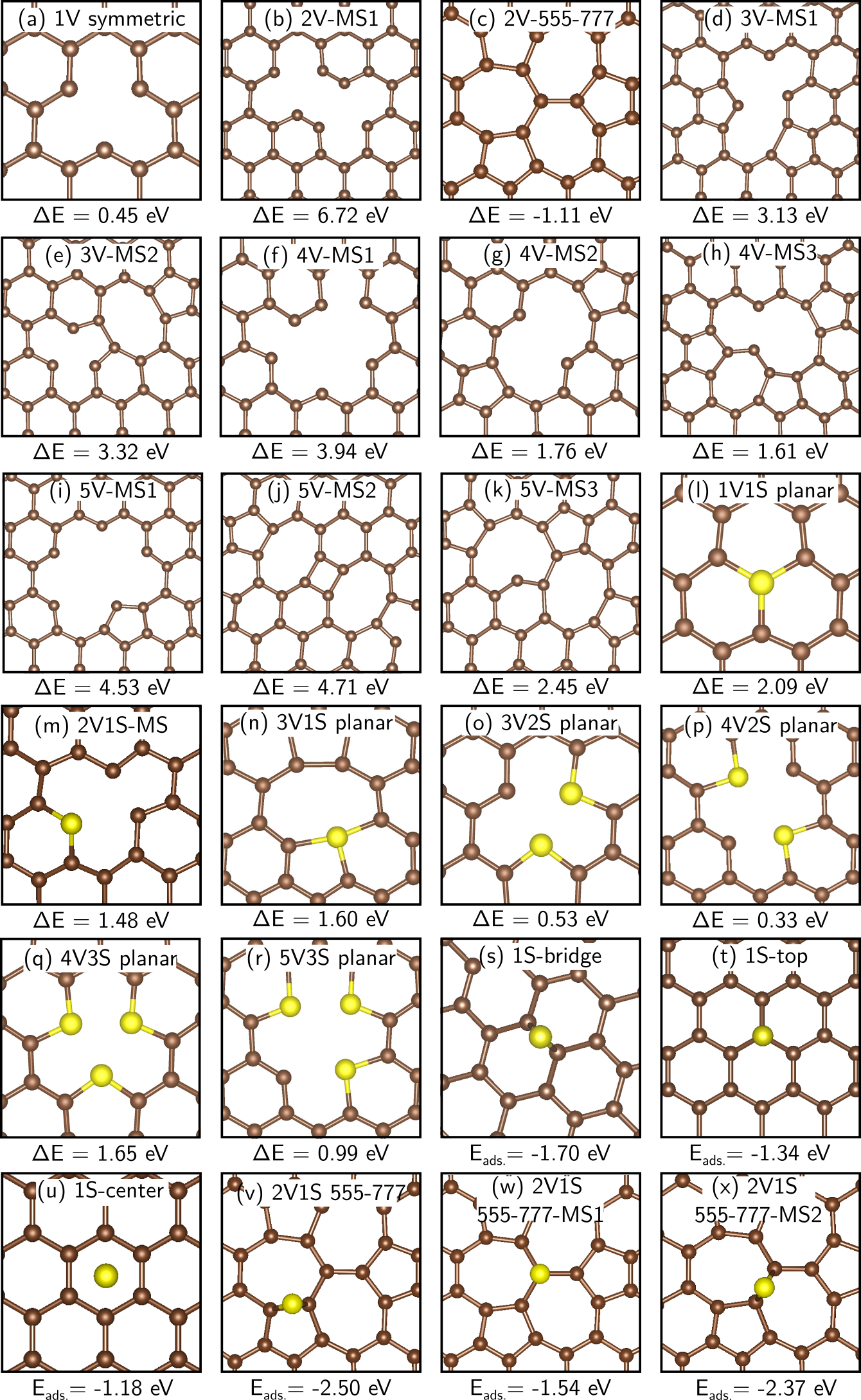}
    \caption{(a-k): Metastable 1V-5V defect structures. In (l-r) top views of additional metastable S-doping configurations are shown. The relative energy to the lowest energy configuration presented in the main text is given below each figure. (s-x): Configurations of S adsorbed on pristine graphene and the reconstructed 2V 555-777 defect and the corresponding adsorption energies. 
    }
    \label{fig:meta-stable}
\end{figure} 

The single vacancy with threefold symmetry has three dangling bonds, as shown in Fig.~\ref{fig:meta-stable}a 
and is 0.45 eV higher in energy than the reconstructed 1V shown in Fig.~\ref{fig:structure_vacancy}a. 
The metastable 2V-MS1 structure created by removing two C-atoms at opposite sides of the hexagon (cf. Fig.~\ref{fig:meta-stable}b). 
has six unsaturated-bonds, making this structure 6.72 eV less stable compared to the 2V 5-8-5 defect (cf. Fig.~\ref{fig:structure_vacancy}a). 
Electron beam irradiation  can lead to a reconstruction to 2V 555-777 via bond-rotation~\cite{Kotakoski_point_defects_2011} (cf. Fig.~\ref{fig:meta-stable}c). 
While this configuration has a lower formation energy of 5.54~eV, the 
transformation requires a high kinetic barrier~\cite{Kim_2011_divacancy_kinetic_barrier}.  
Moreover, the reconstructed 2V is less chemically active, e.g., in binding dopands due to its closed ring structure 
~\cite{defect_reactivity_graphene_2013}. 
We found for example that the S-dopant could not adsorb into the 5- or 7-ring of graphene and instead bonded to a bridge position between two C-atoms of the 5-ring. For these reasons this configuration was not further considered

For 3V we have considered additionally 3V-MS1 and 3V-MS2, shown in Fig.~\ref{fig:meta-stable}d and e, respectively. 
The 3V-MS1 structure, obtained by removing two C-atoms at opposite sides of a hexagon and an additional neighboring C-atom, has three dangling bonds and is 3.13~eV less favorable than the 3V configuration with one dangling bond (cf. Fig.~\ref{fig:structure_vacancy}). 
The 3V-MS2 structure consists of a 1V and a 2V defect next to each other with a total of three dangling bonds, 3.32~eV higher in energy compared to the 3V configuration (cf. Fig.~\ref{fig:structure_vacancy}a). This shows that the extended defect in the 3V configuration is more stable than neighboring 1V and 2V defects. 

The ground state 4V structure (cf. Fig.~\ref{fig:structure_vacancy}a) has no unsaturated bonds and  is 3.94~eV, 1.76~eV and 1.61~eV more favorable compared to the metastable 4V-MS1, 4V-MS2 and 4V-MS3, presented in Fig.~\ref{fig:meta-stable}f-h, which have six (4V-MS1) and two (4V-MS2 and 4V-MS3) dangling bonds, respectively. 4V-MS1 is created by removing four consecutive C-atoms of a hexagon, while 4V-MS2 is created by removing two neighboring C-atoms at each side of the hexagon. Notably, the 4V-MS3 structure (cf. Fig.~\ref{fig:meta-stable}h)
consists of a 7-ring and an almost closed 8-ring in proximity to each other, created by combining two neighboring 2V defects and presents two dangling bonds.

For the 5V defect, metastable structures are shown in Fig.~\ref{fig:meta-stable}i-k. 5V-MS1 is created by removing five C of a hexagon and has five dangling bonds in total, compared to one in 5V  (cf. Fig.~\ref{fig:structure_vacancy}a). Thus, 5V-MS1 is associated with an energy increase of 4.53~eV compared to 5V (cf. Fig.~\ref{fig:structure_vacancy}a). 5V-MS2 consists of two neighboring 2V 5-8-5 defects and has no dangling bonds. However, the formation of a 4-ring at the contact point between the two 2V defects is unfavorable compared to 5- or 6-rings, thus associated with 4.71~eV increase in energy compared to 5V (cf. Fig.~\ref{fig:structure_vacancy}a). 5V-MS3 has neighboring 8- and 9-ring with one dangling bond in the latter. The total energy is 2.45~eV higher compared to 5V (cf. Fig.~\ref{fig:structure_vacancy}) and shows that an extended 5V defect is more energetically favorable than two separated vacancies. 

Additionally, we also considered different surface adsorption configurations of the S-dopant on  pristine graphene. 
The adsorption energy is defined as:
\begin{equation}
\label{eq:adsorbtion}
    E_\text{ads} = (E_{\text{S:G}}-E_\text{G}-E_\text{S-atom})
\end{equation}
where $E_{\text{G}}$ and $E_\text{S:G}$ represent the total energy of graphene and graphene with an adsorbed S, respectively. $E_\text{S-atom}$ represents the total energy of a S-atom.  
The preferred adsorption configuration of S on pristine graphene is a bridge position between two C-atoms  with $E_\text{ads} = -1.70$~eV (cf. Fig.~\ref{fig:meta-stable}s). An S-atom on the top of a C-atom (cf. Fig.~\ref{fig:meta-stable}t) is less favorable  ($E_\text{ads} = -1.34$~eV). An S-atom placed in the center of a 6-ring is associated with an adsorption energy of -1.18 eV and is the least favorable position (cf. Fig.~\ref{fig:meta-stable}u). 

Due to the 2V 555-777 configuration not presenting dangling bonds, an S-incorporation into the lattice structure of graphene was not found and the surface adsorption was considered instead.   
The most favorable adsorption site with $E_\text{ads} = -2.50$~eV is at the bridge position between two C-atoms of a 5-ring (cf. Fig.~\ref{fig:meta-stable}v), followed by adsorption in the bridge position between two C-atoms of a 7-ring (-2.36~eV) (cf. Fig.~\ref{fig:meta-stable}x) and at an on-top position at the central C-atom between the three 7-rings ( -1.51~eV ) (cf. Fig.~\ref{fig:meta-stable}w).
We find that the 2V 555-777 configuration leads to an overall stronger binding of the S-dopant compared to pristine graphene.

\bibliography{Bibtex/S-FLG,Bibtex/References}

\end{document}